\documentclass[aps,amsmath, prl,twocolumn,superscriptaddress,floatfix,amsfonts]{revtex4-2}

\usepackage[english]{babel}
\usepackage[utf8]{inputenc}
\usepackage{xr-hyper}

\usepackage[colorlinks]{hyperref}
\usepackage[T1]{fontenc}
\usepackage[dvips]{graphicx}
\usepackage{amsmath}
\usepackage{amsfonts}
\usepackage{color}
\usepackage{ulem}
\usepackage[caption=false, justification=justified]{subfig}
\usepackage{url}
\usepackage{ulem}
\usepackage[top=2cm, bottom=2cm, left=2cm, right=2cm]{geometry}

\makeatletter
\newcommand*{\addFileDependency}[1]{% argument=file name and extension
 \typeout{(#1)}
 \@addtofilelist{#1}
 \IfFileExists{#1}{}{\typeout{No file #1.}}
}
\makeatother

\usepackage{xcolor}

\hypersetup{filecolor=blue}
\hypersetup{citecolor=blue}
\hypersetup{urlcolor=blue} 

\hypersetup{linkcolor=blue}

\begin{document}

\title{Probing nonlinear spin dynamics in canted easy-plane antiferromagnets using spin-rectification effects}
%Including the Authors/preliminary order 
\author{A. El Kanj}
\affiliation{Laboratoire Albert Fert, CNRS, Thales, Université Paris-Saclay, 91767 Palaiseau, France}
\author{S. Mantion}
\affiliation{Laboratoire Albert Fert, CNRS, Thales, Université Paris-Saclay, 91767 Palaiseau, France}
\author{I. Boventer}
\affiliation{Laboratoire Albert Fert, CNRS, Thales, Université Paris-Saclay, 91767 Palaiseau, France}
\author{P. Bortolotti}
\affiliation{Laboratoire Albert Fert, CNRS, Thales, Université Paris-Saclay, 91767 Palaiseau, France}
\author{V. Cros}
\affiliation{Laboratoire Albert Fert, CNRS, Thales, Université Paris-Saclay, 91767 Palaiseau, France}
\author{A. Anane}
\affiliation{Laboratoire Albert Fert, CNRS, Thales, Université Paris-Saclay, 91767 Palaiseau, France}
\author{O. Gomonay}
\affiliation{Institut f\"{u}r Physik, Johannes Gutenberg-Universit\"{a}t Mainz, Staudingerweg 7, D-55099 Mainz, Germany}
\author{R. Lebrun}
\affiliation{Laboratoire Albert Fert, CNRS, Thales, Université Paris-Saclay, 91767 Palaiseau, France}

\begin{abstract} 
We investigate spin-rectification phenomena in canted antiferromagnets, closely connected to the family of altermagnetic materials. Our results show that excitation efficiency is significantly enhanced by the Dzyaloshinskii-Moriya interaction. Antiferromagnetic dynamics can be detected through spin-Hall magnetoresistance and bolometric effects, with an efficiency reaching up to mV/W. The rectified voltage shape is influenced by both the symmetry of the exciting torques, detection mechanisms (continuous spin-pumping and spin-Hall magnetoresistance), and the antiferromagnetic crystalline axis. Under high pumping power, we observe a saturation effect related to Suhl-like spin-wave instabilities and a nonlinear redshift of the antiferromagnetic resonance. These findings open new avenues for studying nonlinear dynamics in antiferromagnetic and altermagnetic spintronic devices.
\end{abstract}
%\keywords{asfasfadsfadsfadsf}
\maketitle
%\newpage
%%%%%%%%%%%%%%%%%%%%%%%%%%%%%%%%%%%%%%%Manuscript%%%%%%%%%%%%%%%%%%%%%%%%%%%%%%%%%%%%%%%%%%%%
%\section{Introduction}
Spin dynamics in antiferromagnets (AFMs) and altermagnets offer exciting prospects for spintronic devices operating at frequencies beyond several GHz, reaching into the THz range \cite{baltz_antiferromagnetic_2018, kampfrath_coherent_2011, rongione_emission_2023}. AFMs also provide access to rich, largely unexplored physics, such as inertial spin dynamics \cite{kimel_inertia-driven_2009}, eigenmodes with noncircular polarization \cite{johansen_spin_2017}, nonlinear dynamics between eigenmodes \cite{kamra_antiferromagnetic_2019, leenders_canted_2024}, and superfluid states \cite{takei_superfluid_2014}. However, the vanishing net magnetic moment, combined with exchange-enhanced linewidths, complicates the detection and control of AFM magnetization dynamics using conventional electronic methods. In contrast, altermagnets\cite{smejkal_emerging_2022}, and the closely related family of canted antiferromagnets\cite{leenders_canted_2024}, inherit the ultra-fast AFM responses and the presence of small net magnetization in their AFM domains or domain walls however enables an easier control of AFM domains by external magnetic fields and magnetic field gradients \cite{gomonay_structure_2024}.
To date, magnonic studies in antiferromagnets have primarily focused on uniform oscillations \cite{elliston_antiferromagnetic_1968}, with only recent reports of spin-wave dynamics in single crystals \cite{el_kanj_antiferromagnetic_2023, hamdi_spin_2023, wang_long-distance_2023}. In parallel, the advent of high-power femtosecond lasers has enabled the first observations of nonlinear uniform spin-excitations in antiferromagnets \cite{leenders_canted_2024, mukai_nonlinear_2016, yang_spin-orbit_2024, bossini_ultrafast_2021}. Spintronic-based approaches for electrically accessing and detecting AFM spin dynamics have also been proposed. \cite{cheng_spin_2014, cheng_terahertz_2016, khymyn_antiferromagnetic_2017}. 

In ferromagnets, spin-dynamics are commonly detected via spin-diode effects, in which injected microwave currents are rectified through magnetoresistive oscillations or spin-pumping-induced continuous voltages \cite{chiba_current-induced_2014}. The shape of the rectified voltage depends on the type of excitation (e.g., spin torques or current driven-fields) and the detection mechanism (e.g, spin-pumping or spin-Hall magnetoresistance) \cite{jungfleisch_insulating_2017, jungfleisch_large_2016}. An alternative bolometric detection method, using an adjacent platinum thermal sensing layer, is also possible \cite{gui_electrical_2007}. These rectification mechanisms could similarly be applied to study AFM spin dynamics. However, spin-diode phenomena in antiferromagnets have been theoretically explored in collinear AFMs for spin-orbit torque excitation \cite{johansen_spin-transfer_2018, khymyn_antiferromagnetic_2017}, and lacks of experimental evidence, aside from coupled dynamics between ferromagnets and AFMs \cite{zhou_spin-torquedriven_2024}. Theoretically, detection efficiencies as large as mV/mW can be anticipated \cite{khymyn_antiferromagnetic_2017}, which surpass standard inverse spin-Hall measurements by several order of magnitudes \cite{boventer_room-temperature_2021, wang_spin_2021}. In microstrip geometries with a small cross-sectional area, the generation of strong microwave fields near the platinum detecting interface could enable the study of nonlinear AFM spin dynamics. At high excitation powers, this approach could extend the study of Suhl instability concepts—originally developed for ferromagnets, involving coupling between $k=0$ and $k \neq 0$ modes—to antiferromagnetic materials \cite{suhl_theory_1957, heeger_spin-wave_1963, cole_antiferromagnetic_1967, ozhogin_nonlinear_1965}. Canted antiferromagnets and altermagnets on the other hand, thanks to their non-compensated spin-dynamics\cite{gomonay_structure_2024}, offer the ideal playground to access electrically spin diode effects and nonlinear AFM spin dynamics.

In this article, we demonstrate spindiode and bolometric detection of antiferromagnetic resonance in both linear and nonlinear regimes using bilayers of the canted easy-plane antiferromagnet $\alpha$-Fe$_{2}$O$_{3}$ and a heavy metal. By combining experiments with analytical models, we identify contributions to the rectified voltage from spin-Hall magnetoresistance and spin-pumping, revealing that the shape of the rectified voltage is strongly dependent on the orientation of the antiferromagnetic crystalline axis. Additionally, we show that detection efficiency can be enhanced by a continuous current through bolometric detection of the antiferromagnetic resonance in the heavy metal. At low power, we measure a spin-rectified voltage with a record efficiency of 1 mV/W, comparable to that of ferromagnets, attributed to the large spin-Hall magnetoresistance \cite{fischer_spin_2018, lebrun_anisotropies_2019}. At higher pumping power, we observe the onset of a spin-wave instability, resulting in linewidth broadening and saturation of the rectified voltage.

\begin{figure}[t]
\centering
\includegraphics[width=1\linewidth]{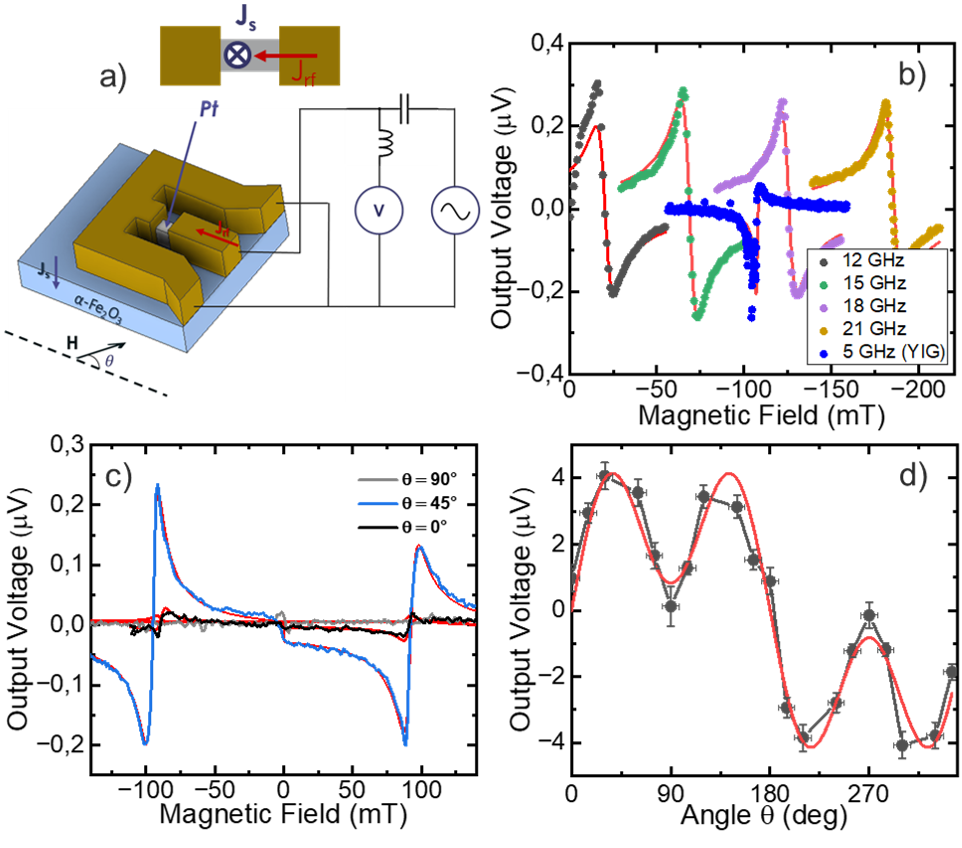}
\caption{\label{Fig_1}  Spin-rectification in 10 $\mu$m wide stripes of bilayers of the canted antiferromagnet $\alpha$-Fe$_{2}$O$_{3}$ (with c-plane orientation) capped with 5 nm of platinum. (a) Sketch of the experimental devices. The radiofrequency excitation and detection are respectively performed through a bias-tee by a microwave source (1-22 GHz) and by a lock-in amplifier. An external magnetic field is applied at angle $\theta$ from the direction of the platinum stripes. (b) Spin-rectification for different exciting frequencies at $\theta$= 45$^\circ$ (c) Exemplary spectra for $\theta$ =0$^\circ$, 45$^\circ$ and 90$^\circ$ at 17 GHz. The small asymmetry arises from nonfully saturated states when sweeping from negative to positive magnetic field. (d) Angular dependency of the rectified voltage. Red line shows the expected theoretical dependencies from Eq.~\eqref{eq_Vsp_cplane}.} 
\end{figure}

We first investigate antiferromagnetic spin-rectification by injecting a resonant radiofrequency current into a 5 nm thick platinum stripe deposited on c-plane oriented single crystals of $\alpha$-$Fe_{2}O_{3}$ in the canted easy-plane AFM phase (see Fig.~\ref{Fig_1}a). At the AFM resonance, we successfully detect a rectified continuous voltage in the microvolt range using a lock-in amplifier, which is approximately two orders of magnitude larger than the inverse spin-Hall signal measured at similar input power \cite{boventer_room-temperature_2021,wang_spin_2021}. As shown in Fig. \ref{Fig_1}.b, the frequency dependence of the rectified voltage aligns with the dispersion relation of the low-frequency mode of hematite \cite{elliston_antiferromagnetic_1968,boventer_room-temperature_2021}. While the amplitude of the rectified voltage is comparable to that of a similar device patterned on a 1 µm thick YIG film, its shape is notably asymmetric, showing a pronounced anti-symmetry with respect to the resonance field. Additionally, the sign of the rectified voltage is opposite for $\alpha$-$Fe_{2}O_{3}$ compared to YIG. As shown in Fig.~\ref{Fig_1}.c, we also observe that the sign of the rectified voltage reverses with the opposite direction of the applied magnetic field, consistent with spindiode phenomena \cite{chiba_current-induced_2014}. In Fig.~\ref{Fig_1}.d, we present the angular dependence of the rectified voltage, revealing $\pi$ periodic oscillations that reach their maximum when the magnetic field is applied at 45$^\circ$ relative to the direction of the injected RF current. These two features (the angular dependence and the anti-symmetric shape) differ from those observed in antiferromagnetic spin-pumping measurements \cite{li_spin_2020, vaidya_subterahertz_2020}.

To analyze the origin of this rectified voltage, we develop an analytical model based on the approach by Chiba et al. for ferromagnets \cite{chiba_current-induced_2014}. We first derive the expressions for magnetization dynamics in the presence of both Oersted field and spin-torque excitation (see Supplementary \cite{noauthor_supplementary_nodate} for details):
\begin{eqnarray}\label{eq_dynamics_general}
&&\mathbf{n}\times\left[\ddot{\mathbf{n}}+\Delta \omega\dot{\mathbf{n}}-\gamma^2 H_\mathrm{ex}\mathbf{H}_\mathbf{n}\right]=\\
&=&    \gamma^2 H_\mathrm{DMI} \left(\hat{z}\times \mathbf{n}\right)\times \mathbf{h}(I(t))-\gamma^2H_\mathrm{ex}\Lambda{\mathbf{n}\times\left[\mathbf{n}\times\boldsymbol\mu_0(I(t))\right]}\nonumber\\
&-&\gamma \mathbf{n}\times \left[\dot{\mathbf{h}}(J(t))-\Lambda\mathbf{n}\times\dot{\boldsymbol\mu_0}(J(t))\right]\times \mathbf{n}.\nonumber
\end{eqnarray}
%$\Delta \omega=\frac{1}{2}\alpha_G\gamma H_\mathrm{ex}$

%\begin{eqnarray}\label{eq_LLG_neel}
 % \dot{\mathbf{m}} &=&-\gamma\left[\mathbf{m}\times{\color{magenta}\mathbf{H}_\mathbf{m}}+\mathbf{n}\times\mathbf{H}_\mathbf{n} \right]+\frac{1}{2}\alpha_G\left[\mathbf{m}\times\dot{\mathbf{m}}+\mathbf{n}\times\dot{\mathbf{n}}\right]\nonumber\\
  %&+&\gamma \Lambda \left[{\color{DarkRed}\mathbf{n}\times\left(\mathbf{n}\times\boldsymbol\mu_0(t)\right)}+\mathbf{m}\times{\color{magenta}\left(\mathbf{m}\times\boldsymbol\mu_0(t)\right)}\right],\\
  %\dot{\mathbf{n}} &=&-\gamma\left[{\color{blue}\mathbf{m}\times\mathbf{H}_\mathbf{n}}+\mathbf{n}\times{\color{magenta}\mathbf{H}_\mathbf{m}}\right]+{\color{blue}\frac{1}{2}\alpha_G\left[\mathbf{m}\times\dot{\mathbf{n}}+\mathbf{n}\times\dot{\mathbf{m}}\right]}\nonumber\\
  %&+&\gamma \Lambda \left[{\color{blue}\mathbf{m}\times\left(\mathbf{n}\times\boldsymbol\mu_0(t)\right)}+\mathbf{n}\times{\color{magenta}\left(\mathbf{m}\times\boldsymbol\mu_0(t)\right)}\right].\nonumber
%\end{eqnarray}
% rewrite with Joe Oersted field and theta*J spin-torque
Here $\mathbf{n}=(\mathbf{M}_1-\mathbf{M}_2)/M_s$ is the N\'eel vector ($|\mathbf{M}_1|=|\mathbf{M}_2|=M_s$), $\mathbf{H}_n$ is the effective anisotropy field, $\gamma$ is the gyromagnetic ratio, $\Delta \omega$ is the linewidth (inverse relaxation time). $H_\mathrm{ex}$ is the intersublattice exchange field, while the DMI field $H_\mathrm{DMI}$ is responsible for small canted moment $\mathbf{m}_0=H_\mathrm{DMI}\left(\hat{z}\times \mathbf{n}\right)/H_\mathrm{ex}$ in equilibrium state. Vectors $\boldsymbol{\mu}_0$ and $\mathbf{h}(t)$ respectively describe the spin accumulation at the Pt interface and the ac magnetic field generated by the ac current $J(t)$. Formally, $\mathbf{h}(t)$ also includes a spin-accumulation contribution (field like torque), proportional to the imaginary spin-mixing conductance $G_i$, which is negligible in spin-Hall magnetoresistance measurements \cite{lebrun_anisotropies_2019,fischer_spin_2018}. Spin-pumping and spin-torque contributions are characterised by the constant 
\[\Lambda=\frac{ G_r}{4e^2M_sd_\mathrm{AF}[1+2\lambda\rho G_r\coth\frac{d_N}{\lambda}]}\]
which depends on the real spin-mixing conductance $G_r$, the electron charge e, the spin diffusion length $\lambda$, the Pt resistivity $\rho$, and the thicknesses of Pt $d_N$ and of the antiferromagnet $d_\mathrm{AF}$. While these spin-torque contributions become negligible in the limit of a thick antiferromagnetic layer $d_\mathrm{AF}\rightarrow \infty$, where $\Lambda\rightarrow 0$, they can be significant in thin films. Notably, the first two terms on the right-hand side of Eq.~\eqref{eq_dynamics_general} describe the current driven ac field and anti-damping torques similar to those in FMs, while the third and fourth terms are specific to AFM and play an important role in the presence of an ac field that we will discuss later.

The generated antiferromagnetic spin dynamics produces a spin current that propagates toward the heavy metal, leading to a continuous voltage through both spin-pumping and the rectification of spin-Hall magnetoresistance (SMR) oscillations. The contributions can be expressed as $V_\mathrm{SMR}\propto\left[n_{0y}\langle J_c^{(0)}(t)\delta n_y\rangle+m_{0y}\langle J_c^{(0)}(t)\delta m_y\rangle\right]$ and $V_\mathrm{SP}\propto\langle \left[\delta\mathbf{n}\times \delta\dot{\mathbf{n}}+\delta\mathbf{m}\times \delta\dot{\mathbf{m}}\right]\cdot\hat{y}\rangle$.
Note that there are two contributions in $V_\mathrm{SMR}$, however, the second one (which is proportional to magnetization) is substantially smaller (as confirmed by the negative sign of the SMR in AFMs as hematite \cite{fischer_spin_2018,lebrun_anisotropies_2019}) and can be neglected. Similarly, in $V_\mathrm{SP}$, when the easy-plane and the sample plane coincides as in c-cut hematite,  the oscillating term of the low frequency mode proportional to the Néel vector leads to out-plane spin-accumulation and can thus be disregarded. We then derive the expressions for the rectified voltage generated at resonance by both $V_\mathrm{SMR}$ and $V_\mathrm{SP}$ (see Supplementary \cite{noauthor_supplementary_nodate} for the details):
\begin{eqnarray}\label{eq_Vsp_cplane}
  % V_{SP}&=&-\frac{4\pi^2\hbar \theta_\mathrm{H}\Delta\rho_1 d_Nh}{ec^2\Delta\rho_0}\left(J_c^{(0)}\right)^2\frac{ H_\mathrm{DMI}\omega_\mathrm{ac}^2}{  H_\mathrm{ex}}F^2(H_0;\omega_\mathrm{ac})\sin2\theta\cos\theta,\nonumber\\
  V_\mathrm{SMR}&=&- \frac{\pi hd_N}{c}\Delta\rho_1 J^{2}F(H_0;\omega_\mathrm{ac})\sin2\theta \cos\theta, \nonumber\\
  V_{SP}&=&\frac{4\pi\hbar \theta_\mathrm{SH}}{ec\Delta\rho_0}\frac{ H_\mathrm{DMI}\omega_\mathrm{ac}^2}{  H_\mathrm{ex}}F(H_0;\omega_\mathrm{ac})V_\mathrm{SMR},
\end{eqnarray}
where the line shape is defined by the function
\begin{equation}\label{eq_line_shape}
    F(H_0;\omega_\mathrm{ac})=\frac{\gamma\left(H_0-H_\mathrm{r}\right)+\delta\Delta\omega\omega_\mathrm{ac}/\gamma H_\mathrm{DMI}}{\gamma^2\left(H_0-H_\mathrm{r}\right)^2+\Delta\omega^2\left(\omega_\mathrm{ac}/\gamma H_\mathrm{DMI}\right)^2}
\end{equation}
Here $H_\mathrm{r}$ is the resonance field \cite{boventer_room-temperature_2021}, $h$ is the length of the Hall bar, $c$ is the speed of light, $\theta_\mathrm{SH}$ is the Pt spin-Hall angle,  $\omega_\mathrm{ac}$ is the ac current frequency, $\hbar$ is the Planck constant. $\Delta\rho_0$ and $\Delta\rho_1$ are the conventional SMR resistivities as introduced in \cite{chen_theory_2013}, and $\delta=\frac{ \Lambda \lambda \theta_\mathrm{SH}\rho ce}{\pi \gamma H_{DMI}} \tanh\frac{d_N}{2\lambda}/d_N$ a prefactor normalizing the Oersted field induced torque and the anti-damping torque (respectively the first and second terms in Eq.~\eqref{eq_dynamics_general}).
%\[\delta=\frac{  \theta_\mathrm{SH}c}{4\pi e d_\mathrm{AF} \gamma H_\mathrm{DMI} M_s}\frac{ \lambda\rho G_r \tanh\frac{d_N}{2\lambda}/d_N }{[1+2\lambda\rho G_r\coth\frac{d_N}{\lambda}]}\]
%\[\Lambda=\frac{ G_r}{4e^2M_sd_\mathrm{AF}[1+2\lambda\rho G_r\coth\frac{d_N}{\lambda}]}\]
%\[\delta=\frac{  \theta_\mathrm{SH}c e \Lambda}{\pi   \gamma H_\mathrm{DMI} }{ \lambda\rho  \tanh\frac{d_N}{2\lambda}/d_N }\]
As mentioned above, $\delta \propto 1/d_\mathrm{AF}$ so that the interfacial contributions (symmetric in shape) from the current induced spin-torque can be neglected for thick films ($>$ 30 - 50 nm). The bulk contributions generated by the Oersted field in $V_{SP}$ and $V_{SMR}$ exhibit the same $\pi$ periodic angular dependence (see Eq.~\ref{eq_Vsp_cplane}) but $V_{SP}$ is symmetric whilst $V_{SMR}$ is anti-symmetric. In c-cut samples, the measured spin-rectified voltage is anti-symmetric (Fig. \ref{Fig_1}b-c), indicating that it mainly arises from the rectification of the injected radiofrequency current combined with oscillations of the spin-Hall magnetoresistance.This observation is also in line with the large spin-Hall magneto-resistance ratio of hematite \cite{fischer_spin_2018, lebrun_anisotropies_2019}. However, the situation could differ for other crystallographic orientations, when the AFM easy-plane and the sample plane are not aligned. In these cases, the Oersted contribution includes a non-zero out-of-plane component, which also excites the Néel vector dynamics (see last term of Eq.\eqref{eq_dynamics_general}).

\begin{figure}[b]
\centering
\includegraphics[width=1\linewidth]{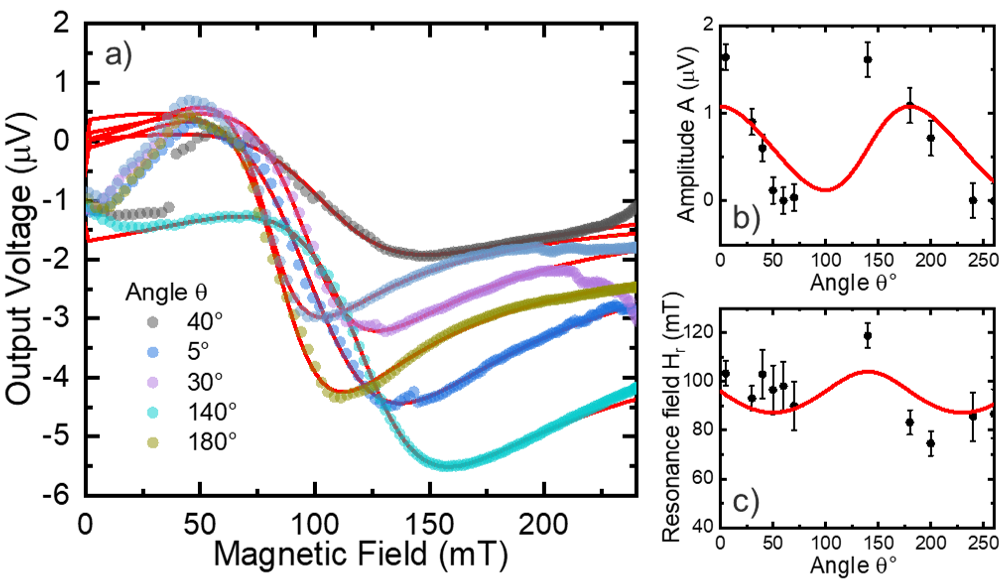}
\caption{\label{Fig_2} Spin-rectification at 17 GHz in 10 $\mu$m wide stripes of bilayers of the canted antiferromagnet $\alpha$-Fe$_{2}$O$_{3}$ (with R-plane orientation) capped with 5 nm of platinum. (a) Examplary spectra for different $\theta$ angles. (b-c) Inplane angular dependency of the resonance field and peak-peak voltage on the applied magnetic field. Solid red lines show the theoretical dependencies calculated according to Eqs.~\eqref{eq_SP_Rplane} and \eqref{eq_resonant field}. }
\end{figure}

To investigate the impact of the cut orientation on the rectified voltage, we examine devices fabricated on a R-plane oriented crystal of hematite, where the easy plane is tilted at $\psi=33^\circ$ from the sample surface. In Fig. \ref{Fig_2}, we present the rectified voltage at 17 GHz in a device oriented at 45$^\circ$ from the in-plane axis of the easy plane \cite{noauthor_supplementary_nodate} for different magnetic fields applied in the device plane. We observe that the shape of the rectified voltage is no longer purely anti-symmetric and varies with the direction of the applied fields. As shown in Fig. \ref{Fig_2}b-c, the angular dependence exhibits a maximum in amplitude around $\theta$=0$^\circ$ and a maximum around 90$^\circ$, shifted by 45$^\circ$ compared to c-plane orientation, and a minimum in resonance field at 45$^\circ$ which belongs to the easy-plane. According to Eq. \eqref{eq_Vsp_cplane}, a symmetric contribution to the rectified voltage can only arise from the presence of an anti-damping torque, which is not expected for thicknesses exceeding 50 nm.

To explain this discrepancy, we extend our modeling approach for arbitrary easy-plane orientation. We derive the revised expressions for the continuous spin-pumping and spin-Hall magnetoresistance voltages, focusing here solely on the contributions from the Oersted field:
\begin{eqnarray}\label{eq_SP_Rplane}
  V_\mathrm{SMR}&=&-\frac{\pi hd_N}{2c}\Delta\rho_1 J^{2}\frac{\sin2\theta F(H_\mathrm{r}^\mathrm{R},\delta_\mathrm{SMR})\cos^3\psi}{1-\sin^2\psi\sin^2\theta}\nonumber\\
  V_{SP}&=&-\frac{4\pi^2\hbar \theta_\mathrm{SH}\Delta\rho_1h d_N}{ec^2\Delta\rho_0}J^{2}(H_\mathrm{r}^\mathrm{R},\delta_\mathrm{SP})\sin^2\psi
\end{eqnarray}
where $F(H_\mathrm{r},\delta_\mathrm{SMR},\omega_\mathrm{ac})$ are defined in Eq.\eqref{eq_line_shape} with the angular-dependent resonant field
\begin{equation}\label{eq_resonant field}
    H_\mathrm{r}^\mathrm{R}=\frac{H_\mathrm{r}^\mathrm{c}}{\sqrt{1-\sin^2\psi\sin^2\theta}},
\end{equation}
and the symmetric prefactors: 
\begin{equation}\label{eq_phase_shift_2}
    \delta_\mathrm{SMR}=\frac{\omega_\mathrm{ac}\tan\psi}{\gamma H_\mathrm{DMI}\cos\theta}\sqrt{1-\sin^2\psi\sin^2\theta}, \, \delta_\mathrm{SP}=0.
\end{equation}
Here $H_\mathrm{r}^\mathrm{c}$ is the resonant field at a given frequency for c-cut.
 %Remarkable, the resonant magnetic field in this case is also angular dependent: $$, where  (see solid line in Fig.~\ref{Fig_3}b). 
%where the symmetric, $F_S$, and antisymmetric, $F_A$, line shapes are
%\begin{eqnarray}\label{eq_R_lineshapes}
%    F_S(H_0,\omega_\mathrm{ac})&=&\frac{\Delta\omega\omega^2_\mathrm{ac}}{\gamma^4H^2_\mathrm{DMI}\left(H_0-H_\mathrm{res}\right)^2+\Delta\omega^2\omega_\mathrm{ac}^2},\nonumber\\
    %F_A(H_0,\omega_\mathrm{ac})&=&\frac{\gamma^2H_\mathrm{DMI}\left(H_0-H_\mathrm{res}\right)\omega_\mathrm{ac}}%{\gamma^4H^2_\mathrm{DMI}\left(H_0-H_\mathrm{res}\right)^2+\Delta\omega^2\omega_\mathrm{ac}^2}.
%\end{eqnarray}

Noteworthy, Eq.~\eqref{eq_SP_Rplane} predicts different angular dependencies for the $V_{SMR}$ and $V_{SP}$  contributions. The $V_{SP}$ contribution is completely anti-symmetric ($\delta_\mathrm{SP}=0$), while the symmetry of the $V_{SMR}$ contribution varies with  $\theta$, allowing it to have a symmetric shape even in the absence of anti-damping torque. Unlike in ferromagnets \cite{chiba_current-induced_2014}, the presence of a sizeable contribution from anti-damping torque cannot be infered solely from the voltage shape. By fitting the experimental angular dependence (see Fig. \ref{Fig_2}c) with Eq. \eqref{eq_SP_Rplane}, we find that $V_{SMR}$ contributes the most to the signal, even for R-plane samples. The amplitude of the rectified voltage remains in the microvolt range for both orientations, similar to that observed in ferromagnets. This aligns with the SMR ratio, which is also on the order of 10$^{-4}$ in both insulating AFMs \cite{ji_negative_2018,fischer_spin_2018,lebrun_anisotropies_2019} and FMs \cite{althammer_quantitative_2013,nakayama_spin_2013}, resulting in comparable detection efficiencies \cite{johansen_spin-transfer_2018}.
%Can you fit with one set of parameters most of the curves ?

\begin{figure}
\centering
\includegraphics[width=1\linewidth]{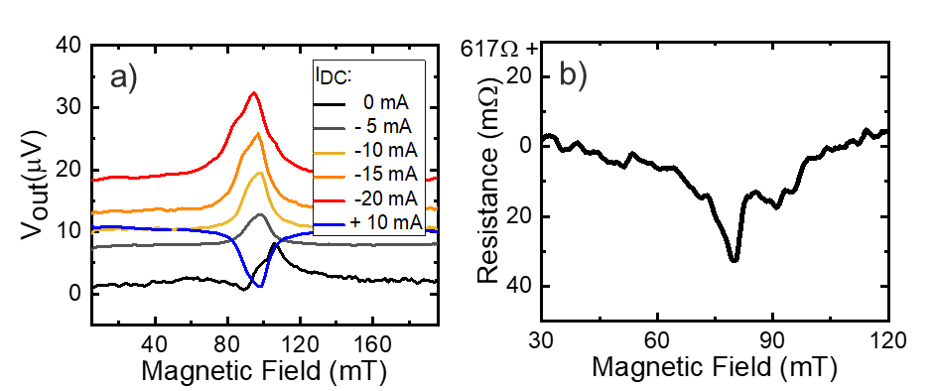}
\caption{\label{Fig_3} Bolometric detection of the AFM resonance (a) Rectified voltage as a function of the applied DC current for a fixed RF power of 3 mW in a 2 $\mu$m wide device. (b) Resistance change in the platinum stripe at resonance for a RF power of 10 mW.}
\end{figure}

We then explore if a continuous (dc) current could modify the spin-rectified voltage even in the case of a bulk AFM crystal. In the simpler case of c-plane orientation (see Fig. \ref{Fig_1}a), the shape of the spin-rectified voltages, shown in Fig. \ref{Fig_3}a, progressively changes from antisymmetric to symmetric when increasing the current from 0 to + 20 mA. This change might appear as a signature of the presence of spin-torque phenomena at the antiferromagnet/heavy metal interface, as indicated by Eq. (\ref{eq_Vsp_cplane}). Notably, we detect a linear increase in voltage with applied dc current, although we do not expect significant spin-orbit torque contributions in micrometer-thick single crystals. However, we also find that the rectified voltage reverses for negative current, which is thus indicative of bolometric detection \cite{gui_electrical_2007}, linked to a temperature rise in the Pt from energy dissipation at antiferromagnetic resonance. To confirm this observation, we measure the resistance of the platinum stripes during resonance excitation, finding a change of about 20 m$\Omega$ at 10 mW (see Fig. \ref{Fig_3}b), corresponding to a temperature increase of roughly 30 K. Thus, we can enhance the detected voltage by increasing the applied dc current. For instance, we achieve in Fig. \ref{Fig_3}a a rectified voltage exceeding 20 $\mu$V at 10 mW with a 20 mA DC current, resulting in an efficiency of 1 mV/W. Using more resistive, for example by narrowing the platinum stripes, we anticipate to be able to reach voltages up to several millivolts. This makes this approach promising for detecting sub-THz signals in other insulating antiferromagnets, with limitations only related to thermal response times (tens of kHz) and the absence of phase resolution. 

\begin{figure}
\centering
\includegraphics[width=1\linewidth]{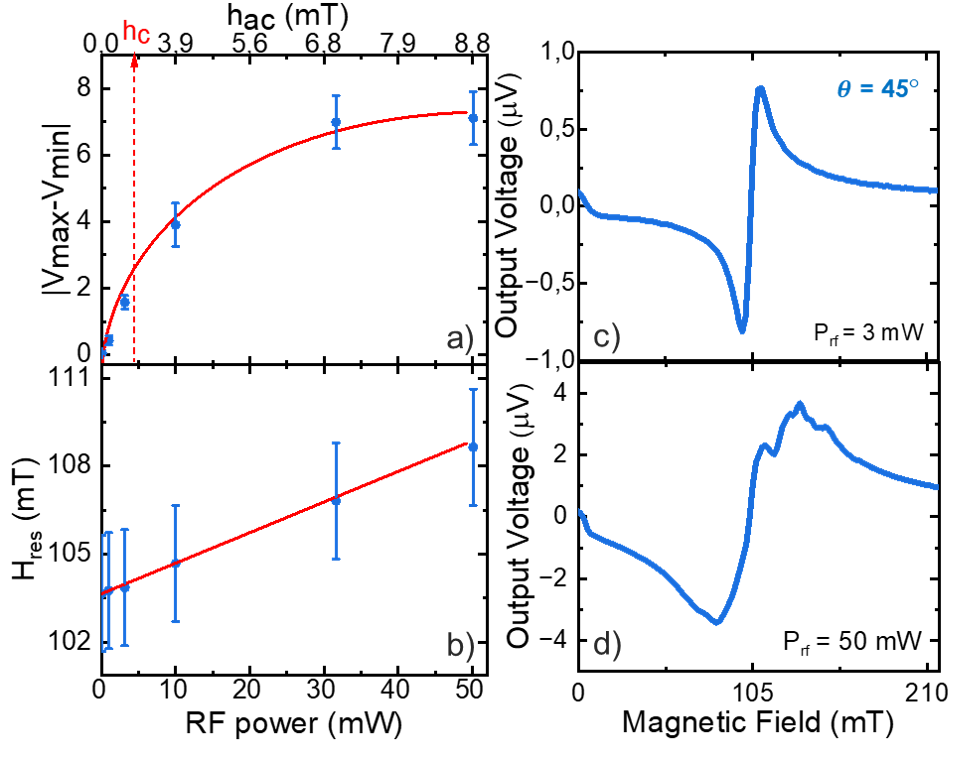}
\caption{\label{Fig_4}  Power dependency of the rectified voltage for a fixed input frequency of 17 GHz in a 10 $\mu$m wide device at $\theta$ = 45$^\circ$. Amplitude (a) and resonance frequency (b) of the rectified voltage with the power of the injected RF current. Red lines corresponds to the theoretical response including nonlinear spin-wave excitation. (c-d) Exemplary spectra at 5 and 17 dBm, showing the onset of a nonlinear behaviour}
\end{figure}

To finally gain further insights into the magnetization dynamics of the antiferromagnet, we record the rectified voltage for varying input RF power (without dc current). As shown in Fig.~\ref{Fig_4}, we observe a linear increase in the rectified voltage up to 31 mW (15 dBm), peaking at around 7 $\mu$V. We also observe that the resonance frequency shifts from 105 to 110 mT. This nonlinear frequency shift arises from large amplitude of uniform oscillations, characterized by a reduction of the averaged Néel vector and consequently of the effective anisotropy. This reduction leads to an amplitude-dependent resonance frequency, which can be linearized for small excitations $\omega_\mathrm{res}\approx\omega_0 (1-A^2/4)$ (see Fig.~\ref{Fig_4}b), where $\omega_0 (H)$ is the AFM resonance frequency, and $A$ is the linearized amplitude of the oscillations, $A\approx \gamma^2H_\mathrm{DMI}h_{0y}/(4\Delta\omega\omega_0)$.

Above 15 mW, we detect a saturation of the rectified voltage alongside linewidth broadening. Firstly, the voltage saturation can be associated with a saturation of the oscillation amplitude. The ac field excites both uniform and nonuniform ($\mathbf{k}\ne0$) oscillations. In the linear regime, the frequency of the nonuniform mode is always higher than that of the uniform mode, and $A$ increases with $h_{y0}$. However, due to nonlinear coupling, the frequency of the nonuniform mode decreases more rapidly than $\omega_\mathrm{res}$. At a critical value $h^\mathrm{cr}_{0y}$, both frequencies coincide, resulting in resonant excitation of the nonuniform mode at high amplitude, which absorbs energy from the pumping field, leading to saturation of the uniform mode's amplitude. This process thus corresponds to a multi-magnon process, corresponding in the conversion from uniform to non-uniform spin-waves mode, and the saturation can be attributed to the Suhl instability \cite{suhl_theory_1957, heeger_spin-wave_1963, cole_antiferromagnetic_1967}. It is noteworthy that the DMI-mediated coupling with the magnetic field plays an important role in the observation of the Suhl instability in easy-plane systems. In particular, the effective ac field depends on the DMI value and on the orientation of the magnetic vectors. Its value decreases as the oscillation amplitude increases, leading to a saturation. 
By adapting Eq.~(1) of Ref. \cite{heeger_spin-wave_1963} to hematite, we thus estimate the critical field for spin-wave instability to be $h^\mathrm{cr}_{0y}\approx (2\Delta\omega)^{3/2}\omega^{1/2}_\mathrm{res}/(\gamma^2H_\mathrm{DMI}) \approx 1$~mT which is exceeded for input power of a few mW \cite{noauthor_supplementary_nodate}. For larger excitation fields, both the uniform mode and the spin-wave mode are excited, contributing to a progressive saturation of the oscillation amplitude \cite{noauthor_supplementary_nodate}. Secondly, the significant linewidth broadening observed at high power, can be attributed to the excitation of nonuniformly propagating modes that transfer energy away from the region beneath the Pt electrode. We can rule out linewidth broadening due to a temperature increase, as the Néel temperature of $\alpha$-Fe$_{2}$O$_{3}$ is approximately 780 K, while the temperature rise in platinum remains below 50 K at 100 mW.

Our findings bring first evidence of the onset of nonlinear effects in AFM spintronic devices and thus demonstrate that spin-rectification effects in AFMs with non-compensated spin dynamics as canted antiferromagnets and altermagnets will unlock unexplored nonlinear AFM dynamics by integrating efficient excitation and detection mechanisms with metallic electrodes. Extending these approaches to thin films could pave the way for efficient sub-THz spin detectors.

The authors acknowledge financial supports from Horizon 2020 Framework Programme of the European Commission under FET-Open grant agreement no. 964931 (TSAR), under Horizon Europe grant agreement no. 101070287 (SWAN) and under the ITN grant agreement ID 861300 (COMRAD). O. G.  acknowledges funding by the DFG Grant No. TRR 288-422213477 (project A12) and TRR 173-268565370 (projects A11 and B15).

%For comparison, in a pure antiferromagnet the magnetic dynamics is excited by a time-dependent component of the magnetic field ($h_z$ in our geometry), which gives the critical field $h^\mathrm{cr}_{0z}\approx (2\Delta\omega)^{3/2}/(\gamma\omega^{1/2}_\mathrm{res})$. For the single plane antiferromagnet $h^\mathrm{cr}_{0y}/h^\mathrm{cr}_{0z}\propto H_\mathrm{sf}/H_\mathrm{DMI}$, where $H_\mathrm{sf}$ is a spin-flop field. In the hematite $H_\mathrm{sf}/H_\mathrm{DMI}\approx 0.03$, demonstrating the efficiency of the DMI mechanism. 
% $ biblatex auxiliary file $
% $ biblatex bbl format version 3.2 $
% Do not modify the above lines!
%
% This is an auxiliary file used by the 'biblatex' package.
% This file may safely be deleted. It will be recreated by
% biber as required.
%


\end{document}